\documentclass[aps,prl,twocolumn,groupedaddress]{revtex4}

\usepackage{graphicx}

\renewcommand{\epsilon}{\varepsilon}

\begin{document}


Published in \textit{Review of Scientific Instruments}, 82, 2011 (in press)
\title{Ice Particles Trapped by Temperature Gradients at mbar Pressure}


\author{Thorben Kelling}
\email[]{thorben.kelling@uni-due.de}
\author{Gerhard Wurm}
\author{Christoph D\"{u}rmann}
\affiliation{Fakult\"{a}t f\"{u}r Physik, Universit\"{a}t Diusburg-Essen, Lotharstrasse 1, 47057 Duisburg}


\date{\today}

\begin{abstract}
In laboratory experiments we observe that ice particles ($\le 100$ $\mu$m) entrained in a low pressure atmosphere ($\sim 1$ mbar) get trapped by temperature gradients between three reservoirs at different tempertature. Confining elements are a peltier element at 250 K (bottom), a  liquid nitrogen reservoir at 77 K (top) and the surrounding vacuum chamber at 293 K. Particle levitation and trapping is modeled by an interplay of thermophoresis, photophoresis and gravity.  A number of ice particles are trapped simultaneously in close spatial distance to each other at least up to minutes and are accessible for further experiments.
\end{abstract}

\keywords{ice particles, thermophoresis, photophoresis}

\maketitle

\section{Introduction}
Ice particles are a major component in the Earth atmosphere and are important for the energy budget and climate \cite{baran2001}. More general, ice is important in planetary atmospheres and in a wider range of astrophysical environments \cite{ehrenfreund2003}. The microphysics of ice particles, e.g. formation or optical properties are regularly studied in large scale simulations \cite{wagner2009}.  Individual ice particles have also been studied trapped in an electrodynamic balance \cite{swanson1999}. An important pressure range for stratospheric and mesospheric particles is the mbar regime. The low pressure range is also required to study photophoresis in detail which has recently gained importance in protoplanetary disks \cite{wurm2009}. This pressure range is intrinsically difficult for Paul traps due to voltage break throughs.

We present here a new kind of particle trap based on thermophoretic, photophoretic and gravitational forces. The trap is capable of levitating ice particles at least up to 100$\mu$m in size at absolute pressures of a few mbar. The trap is static, i.e. does not require alternating fields like for electrodynamic trapping and does not require a feedback loop. Besides trapping particles for further studies (i.e. optical properties) it immediately allows to determine the photophoretic strength on the particles induced by thermal radiation, to determine the aerodynamical behaviour and to quantify thermophoretic forces. The trapped ice particles levitate at least up to minutes at a specific position and are easily accessible. In addition, tens of ice particles can be trapped simultaneously in close spatial distance to each other which allows experiments like inducing collisional interaction. As such we consider this experiment to be not only a tool but to be of fundamental importance for a number of branches in physics.

The experiment is based on forces induced by temperature gradients. The motion of a particle along a temperature gradient in a gaseous environment is called thermophoresis \cite{zheng2002}. Related to this is a force on the particle, where a temperature gradient exists along the particle in a gaseous environment where the gas is at constant temperature. Such a gradient is often induced by radiation, therefore the effect is called photophoresis \cite{rohatschek1995}. Small particles suspended in a rarefied gas with a temperature gradient being present one way or the other are accelerated by a thermophoretic and/or photophoretic force in general in the direction from hot to cold.

Thermophoresis and photophoresis are the result of the interaction between the gas molecules and the particle's surface. Gas molecules colliding with the particle surface, accomodate to the local temperature and leave with a thermal velocity according to the surface temperature. From this a net momentum on the particle results. A detailed theory for thermophoresis and photophoresis exist for spherical particles \cite{zheng2002,rohatschek1995,beresnev1993,beresnev2003}. Many parameters within this theory are not easily accessible for calculations, i.e. the absorption and heat transfer within the particle or the accomodation coefficient are needed which are physical problems of their own. In reality particles are also far away from being perfect spheres with well defined sizes (see also Fig.\ref{fig:levitation}) and correction factors have to be applied. These are currently unknown. If thermophoretic and photophoretic forces are directed upwards they might offset gravity and lead to the levitation of particles. In such a stable case the forces can directly be modelled and the particle can be subjected to further research. To provide stable levitation the lifting phoretic forces have to decrease with height and a sideward confining force is needed to prevent a sideward drift. We also considered convection but convection is too slow in our case and is not important for the trapping principle. However, gas drag is providing damping for a particle trajectory and rotation and allows to determine aerodynamical properties.

\section{Setup and Observations}
\begin{figure}
\centering
\includegraphics[width=0.45\textwidth]{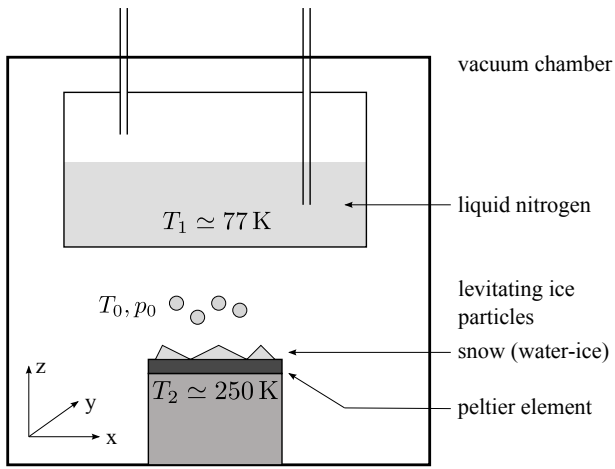}
\caption{Experimental Setup. Snow is placed on a peltier element at $T_2 \simeq 250$ K within a vacuum chamber at room temperature. 
A container filled with liquid nitrogen at $T_1 \simeq 77$ K is $25$ mm above the peltier surface. The chamber is evacuated to $0.1-10$ mbar. Single ice particles ($<100$ $\mu$m) are ejected spontaneously from the snow surface and are trapped between the peltier element and the nitrogen container.}
\label{fig:setup}
\end{figure}

\begin{figure}
\centering
\includegraphics[width=0.45\textwidth]{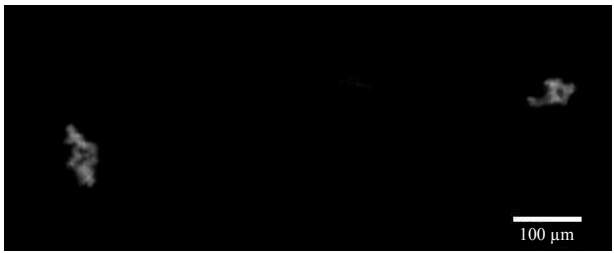}
\caption{Levitating ice particles within the trap between the peltier element and the nitrogen container.}
\label{fig:levitation}
\end{figure}

\begin{figure}
\centering
\includegraphics[width=0.45\textwidth]{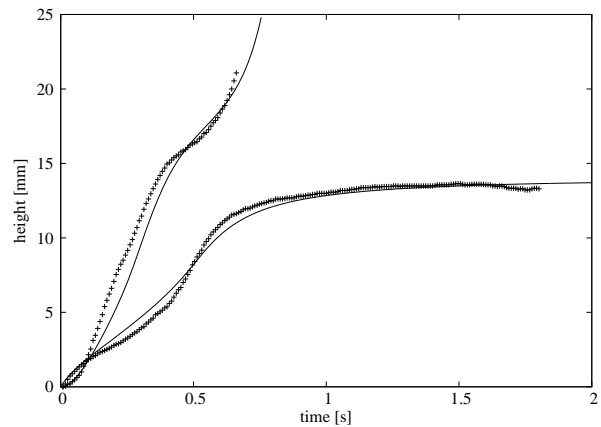}
\caption{Examples of the two main types of trajectories observed for ejected particles in the vertical direction ($p\simeq 1.5$ mbar). Continuous propagation from bottom to top and levitation. Crosses: experimental data; solid lines: model fits (see text).}
\label{fig:heights}
\end{figure}

Fig.\ref{fig:setup} shows the principle setup of the experiment and Fig.\ref{fig:levitation} is an example of levitated ice particles.
An ice sample grown on a cold surface but currently not analyzed further is placed on a peltier element at $T_2 \simeq 250$ K within a vacuum chamber. The cold bottom of a container filled with liquid nitrogen ($T_1 \simeq 77$ K) is located $25$ mm above the peltier element. The chamber is evacuated to pressures between $0.1-10$ mbar. Ice particles are ejected spontaneously from the ice surface at mbar pressure. This might be due to the photophoretic ejection mechanism discussed in \cite{wurm2006} and \cite{kelling2011}. After the particles are ejected, in general two different types of particle trajectories are observed (Fig.\ref{fig:heights}).
Most particles continuously rise with their velocity varying with height. They are decelerated in the lower part of their trajectory but are accelerated once they reach a certain height. The other fraction of particles comes to rest about half the way between bottom and top and is trapped for at least a few minutes. Tens of particles can be trapped at the same time. The sizes of the levitated particles roughly vary between 10 and 100 $\mu$m. The lower limit is set by the resolution of the camera. A minor fraction of particles fall back to the snow surface. Most levitating particles rotate about the vertical axis \cite{eymeren2011}.

\section{Model and Simulations}

\begin{figure}
\centering
\includegraphics[width=0.45\textwidth]{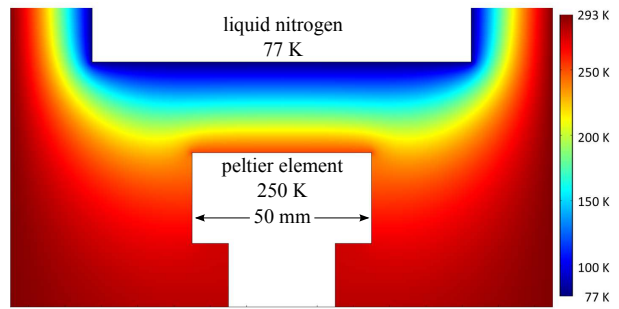}
\caption{Simulation of the temperature distribution within the vacuum chamber.}
\label{fig:temp_dis}
\end{figure}

\begin{figure}
\centering
\includegraphics[width=0.45\textwidth]{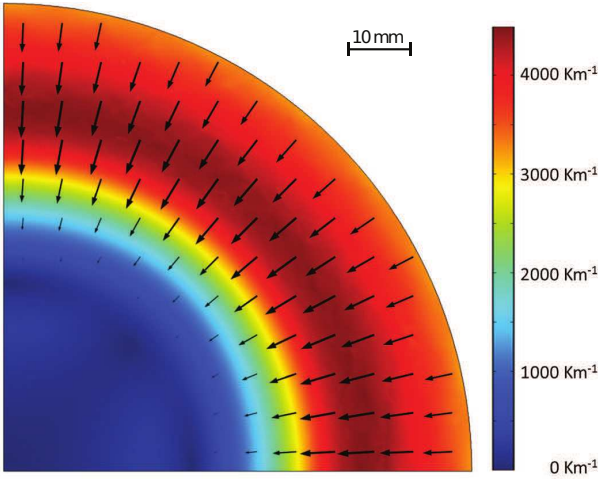}
\caption{Simulation of the horizontal temperature gradient at the trapping height. Levitated particles are dragged towards the center by thermo- and photophoretic forces.}
\label{fig:temp_grad}
\end{figure}

We consider thermophoresis, photophoresis, gravity and gas drag here. While thermophoresis directly depends on the temperature gradient between the peltier element and the nitrogen container, photophoresis is induced by the thermal radiation $I=\sigma T^4$ of the peltier element and -- once rising -- by the surrounding warm chamber ($293$ K). The higher the particle ascends within the gap between the peltier element and the nitrogen container the more dominant is the ambient thermal radiation. Hence photophoresis gets stronger with height $z$. We simulated the gas flow and temperature distribution within the chamber using a commercial software (Comsol Multiphysics 4). Fig.\ref{fig:temp_dis} shows the temperature distribution and Fig.\ref{fig:temp_grad} depicts the temperature gradient from the warm vacuum chamber walls towards the center of the chamber at half the way between the peltier element and the nitrogen container.  

The thermophoretic force for a spherical particle $F_{th}$ is given by \cite{zheng2002}
\begin{equation}
F_{th}=\frac{f_{th}a^2\kappa_g}{\sqrt{2k_B T_0/m_g}}\frac{dT}{dz}\label{eq:thermo}
\end{equation}
with $f_{th} (Kn)$ as dimensionless thermophoretic force depending on $Kn=\lambda/a$, $a$ as the particles radius and $\lambda (T)$ as the temperature dependend mean free path of the gas molecules, $\kappa_g=0.02$ W/mK as the thermal conductivity of the air, $k_B=1.38\times 10^{-23}$ J/K as Boltzmann constant, $T_0 (z)$ as the average temperature at the particles position, $m_g=4.8\times 10^{-26}$ kg as the molecular mass of air and $dT/dz$ as the temperature gradient surrounding the particle. The photophoretic force is \cite{rohatschek1995}
\begin{eqnarray}
F_{ph}&=&\frac{2F_{max}}{\frac{p}{p_{max}}+\frac{p_{max}}{p}}\label{eq:roha}\\
p_{max}&=&\sqrt{\frac{6\pi \kappa}{\alpha}} \frac{\bar{c}\eta}{2a}\\
F_{max}&=&\sqrt{\frac{\alpha\pi\kappa}{6}}\frac{\pi\bar{c}\eta a^2J_1I}{2\kappa_pT}
\end{eqnarray}
with $\kappa=1.14$ as thermal creep coeffcient \cite{rohatschek1995}, $\alpha$ as particle dependent accomodation coefficient, $\bar{c}(T) = \sqrt{(8k_B T)/(\pi m_g)}$ as mean thermal velocity of the gas molecules, $\eta(T)$ as gas dynamical viscosity, $a$ as particle radius, $\left| J_1 \right|=0.5$ as asymmerty factor for a sphere \cite{rohatschek1995}, $p$ as pressure, $I(z)$ as height dependend infalling thermal radiation from the peltier element and the surrounding vacuum chamber, $T(z)$ as height dependend temperature and $\kappa_p$ as the ice particles thermal conductivity. Gas drag at large Knudsen numbers can be written as
\begin{equation}
F_{gas} = c_d\frac{mv}{\tau}\label{eq:gasdrag},
\end{equation}
where $c_d$ is the drag coefficient, $m$ is the particle mass, $v$ is the particles velocity and $\tau$ is the gas-grain coupling time which can be expressed as \cite{blum1996}
\begin{equation}
\tau = \epsilon \frac{m}{\sigma}\frac{1}{\rho_g \bar{c}}\label{eq:tau},
\end{equation}
with an empirical factor $\epsilon$, $m$ as the particles mass, $\sigma$ as the geomoetrical cross section of the particle, $\rho_g$ as the gas density. The gravitational force is $F_G=(4/3) \pi f \rho_p a^3 g$ with $g=9.81$ m/s$^2$ and a filling factor $f=\rho_p/\rho_s$ ($\rho_p$ as the ice particles density and $\rho_s$ as the corresponding density of the solid, non porous material).

\begin{figure}
\centering
\includegraphics[width=0.45\textwidth]{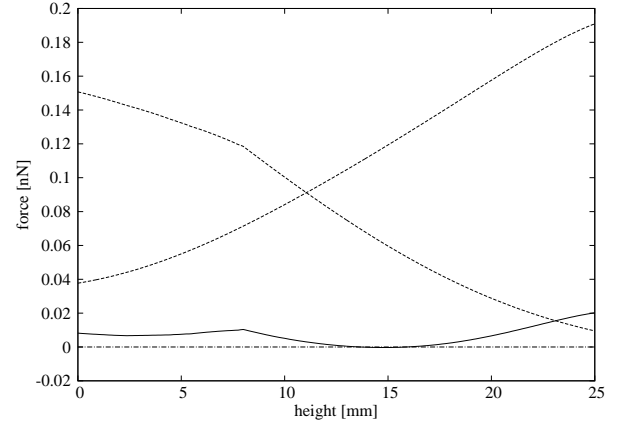}
\caption{Forces acting on the ice particle in the stable levitation case. Solid line: sum of all forces ($F_{ph}+F_{th}+F_G+F_{gas}$), dashed-dotted line marks the zero. The thermophoretic force (decreasing with height) and the photophoretic force (increasing with height) are represented by the dotted lines. Gravity is not depicted - it gives only a constant offset. The gas motion is - at maximum -  two orders of magnitudes lower and is also not shown. Stable levitation should appear in the minimum of the sum of the forces which is in good agreement with the experimental data (see Fig.\ref{fig:heights}).}
\label{fig:sim3}
\end{figure}

At the height $z_0$ where the particles levitate the sum of all forces acting on the particle in $z$-direction must vanish
\begin{equation}
F_{th_z}+F_{ph_z}+F_G+F_{gas_z}=0.\label{eq:zero}
\end{equation}
Fig.\ref{fig:sim3} shows the sum of $F_{th_z}$, $F_{ph_z}$, $F_G$ and $F_{gas_z}$ from our simulations for the case of a levitating ice particle assuming $a\simeq 30$ $\mu$m,  $\rho = 1000$ kg/m$^3$, a filling factor $f=0.13$, a pressure of $p\simeq 1.5$ mbar, a thermal conductivity of the particle $k_p\simeq 4\times 10^{-3}$ W/(m K), $c_d=0.1$, $\alpha=0.34$ and $\epsilon = 0.7$ to fit the data.
In Fig.\ref{fig:heights} the experimental data and the predictions of the theory are plotted. The trajectories are reproduced by the model. For continuous propagation from the bottom to the top the gas drag coefficient takes the value of $c_d = 0.03$. From the model ice particles should levitate approx. at half the distance (minimum of the force curve) between the peltier element and the nitrogen container which is in good agreement with the experiments. Stable trapping is provided here as the $x,y$-components of the thermophoretic force on the ice particles yields from the warm vacuum chamber walls inwards. This can be seen from the temperature gradient (Fig.\ref{fig:temp_grad}) at the levitation height. We note that the observed stable trapping is offset from the center. We attribute this to the the low temperature gradient in the center and peculiarities of the setup, e.g. light sources, flanges and a square peltier element.

\section{Application and Conclusion}
We showed that ice aggreagtes can be trapped and levitated by thermophoresis and photophoresis between two surfaces at different temperatures in a vacuum chamber at room temperature. The levitated ice particles are in general $< 100$ $\mu$m. Even groups of single ice particles can be trapped in close spatial distance. The ice particles allow further experiments like detailed thermophoretic, photophoretic or optical experiments. Also the thermal behaviour of irregular shaped ice particles and the interaction between ice particles e.g. in collisions induced in a trapped cloud can be studied. Possible applications are basic physics on thermophoretic and photophoretic forces and their use in atmospheric science and astrophysics. For the latter we currently explore the trap described here as levitation mechanism in protoplanetary disks. 

\begin{acknowledgments}
We thank Marc Boch for supporting the laboratory work. This work is funded by the Deutsche Forschungsgemeinschaft.
\end{acknowledgments}

\end{document}